\documentstyle[12pt]{article}
\newcommand{\beq}{\begin{equation}}   
\newcommand{\eeq}{\end{equation}}
\newcommand{\ra}{\rightarrow}

\newcommand{\gsim}{\lower.7ex\hbox{$
\;\stackrel{\textstyle>}{\sim}\;$}}
\newcommand{\lsim}{\lower.7ex\hbox{$
\;\stackrel{\textstyle<}{\sim}\;$}}

\begin{document}
\begin{titlepage}
\renewcommand{\thefootnote}{\fnsymbol{footnote}}

\begin{center} \Large
{\bf Theoretical Physics Institute}\\
{\bf University of Minnesota}
\end{center}
\begin{flushright}
TPI-MINN-97/03-T\\
UMN-TH-1530-97\\
hep-th/9702174
\end{flushright}
\vspace*{1.3cm}

\begin{center}
{\Large \bf  Chirally Symmetric Phase of Supersymmetric 
Gluodynamics}
\vspace{0.7cm}

{\Large A. Kovner$^*$ }

\vspace{0.1cm}

{\it  Theoretical Physics, Oxford Univ. 1 Keble Road, Oxford OX13NP, 
UK}

 {\it and}

{\it  Theoretical Physics Institute, Univ. of Minnesota,
Minneapolis, MN 55455}

\vspace{0.5cm}
and
\vspace{0.5cm}

{\Large  M. Shifman} 

\vspace{0.1cm}
{\it  Theoretical Physics Institute, Univ. of Minnesota,
Minneapolis, MN 55455}
\end{center}

\vspace*{.3cm}
\begin{abstract}

We argue that  supersymmetric gluodynamics (theory of gluons 
and 
gluinos) has a condensate-free phase. Unlike the standard phase, 
the discrete axial 
symmetry of the Lagrangian
is unbroken in this phase, and the gluino condensate does not 
develop. Extra unconventional vacua are supersymmetric and are 
characterized by the presence of (bosonic and fermionic) massless 
bound states. A set of arguments in favor of the conjecture includes:
(i) analysis of the effective Lagrangian of the 
Veneziano-Yankielowicz
type which we amend to properly incorporate all symmetries of the 
model; (ii) consideration of an unsolved problem with the Witten 
index; (iii) interpretation of a mismatch between the strong-coupling 
and weak coupling instanton calculations of the gluino condensate 
detected previously. Impact on Seiberg's results is briefly discussed.
 
\end{abstract}

\vspace{1cm}
\begin{flushleft}
$^*$ On leave of absence from PPARC Advanced Fellowship
\end{flushleft}

\end{titlepage}

\section{Introduction}

In this work, supersymmetric gluodynamics,  
 theory of gluons and 
gluinos with no matter, is revisited. Our primary   task is  
investigating the modes of  realization of the discrete chiral 
invariance in supersymmetric gluodynamics. As is 
well-known from the early 
days of supersymmetry, this theory possesses a discrete symmetry
$Z_{2T(G)}$ where $T(G)$ is (one half) of the Dynkin index for the 
given gauge group $G$. In $SU(N)$ supersymmetric gluodynamics 
$T(G) = N$. Since the $Z_{2T(G)}$ invariance is the (non-anomalous) 
remnant of the 
anomalous axial 
symmetry generated by the phase rotations of the gluino field, the 
gluino condensate $\langle\lambda \lambda\rangle$ 
is the order parameter \footnote{The gluino field is treated in the 
Weyl representation; note that the condensate under discussion is 
$\lambda^2$ rather than $\bar\lambda\lambda$.}. Usually it is 
believed that a non-vanishing gluino condensate develops, 
spontaneously breaking $Z_{2T(G)}$ down to $Z_2$. 
Then the space of vacua consists of $T(G)$ points. All these vacua are 
physically equivalent and correspond to confining dynamics 
qualitatively similar to that of non-supersymmetric gluodynamics. In 
particular, a mass gap develops, and massless excitations present at 
the Lagrangian level disappear from the physical spectrum.

The picture seems perfectly self-consistent; yet, two unsolved issues 
have
clouded it for over a decade. First, the number of vacua, $T(G)$,
does not match the  value of the Witten index \cite{WI} for 
orthogonal and exceptional groups. (Say, for the $O(N)$ groups
with even $N$ the index is predicted \cite{WI} to be $(N/2)+1$ while 
$T(G) =N-2$).  Second,
the value of the gluino condensate, calculated in the
weak-coupling regime and analytically continued
to the strong-coupling regime by using holomorphy \cite{SHVA}
does not match $\langle\lambda \lambda\rangle$ calculated directly 
in the strong-coupling regime \cite{NSVZ}. 
More exactly, the direct calculation refers to 
\beq
\langle \lambda\lambda (x) \, \lambda\lambda (0)\rangle\, , \,\,\, 
(G=SU(2))
\eeq
and is carried out {\em via} instantons, plus cluster decomposition
\cite{NSVZ1}. The weak-coupling regime is achieved by adding extra 
matter fields and working in the Higgs phase, with the subsequent 
limit $m\ra\infty$ where $m$ is the matter mass term. 

In this work we suggest a somewhat unexpected solution
which seems to eliminate both difficulties. We will
argue that  extra vacuum states, with unbroken 
$Z_{2T(G)}$ symmetry and vanishing gluino condensate, exist.
The gauge dynamics in this unbroken phase is very peculiar. In 
particular, 
although no symmetry is spontaneously broken,  it should contain 
massless excitations of both bosonic  and fermionic type. 

The above conclusion is based on  two sets of  arguments.
First, the existence of extra vacuum states follows from the 
analysis
of the so-called Veneziano-Yankielowicz (VY) effective 
Lagrangian \cite{VY,TVY}.  A technical problem one 
immediately encounters 
is the absence of $Z_{2T(G)}$
degeneracy in the original VY expression. We show that this 
expression
is incomplete, and explain how it must be amended to become 
compatible with all symmetries of supersymmetric gluodynamics.
The corrected expression exhibits $T(G)$ minima of the scalar 
potential corresponding to $Z_{2T(G)}\ra Z_2$ breaking,
plus  an additional  minimum at the origin
where the gluino condensate vanishes. We then discuss the 
occurrence 
of this extra state
in relation with the Witten-index problem. The $\langle\lambda 
\lambda\rangle = 0$ state presumably does not contribute to the 
Witten index 
counting for the unitary gauge  groups, since it is accompanied by a 
``fermion" zero energy state. It may contribute, however,
in the case of the orthogonal groups. 

Finally, a mismatch between
the direct instanton calculation of the gluino condensate and 
an indirect derivation through the Higgs phase is interpreted
as a signature of the $\langle\lambda \lambda\rangle = 0$
vacuum contribution in the instanton calculation.
A few remarks concerning  infrared dynamics
in the $\langle\lambda \lambda\rangle = 0$ vacuum
and the possible impact of the inclusion of  light matter conclude the 
paper.

\section{Veneziano-Yankielowicz Effective Lagrangian}

In this section we  discuss effective Lagrangians and  the manifold of 
vacua in 
supersymmetric Yang-Mills theory without matter \footnote{This 
theory is referred to as supersymmetric gluodynamics. The theory 
where the light matter fields in the fundamental representation are 
included will be referred to as SUSY QCD.}. The Lagrangian 
of 
the model at the fundamental level  is
\beq
{\cal L} = \frac{1}{g_0^2}\left[ -\frac{1}{4} 
G_{\mu\nu}^aG_{\mu\nu}^a
+ i\lambda_{\dot\alpha}^\dagger D^{\dot\alpha\beta}\lambda_\beta
\right] +{\vartheta\over 32\pi^2}G_{\mu\nu}^a\tilde G_{\mu\nu}^a\, 
,
\label{SUSYML}
\eeq
where it is assumed, for simplicity, that the gauge group $G$ is 
$SU(N)$.
This model possesses a discrete global $Z_{2N}$ symmetry, a 
residual non-anomalous subgroup of the anomalous chiral $U(1)$.  

One of the aspects of our consideration is based  on  the effective 
Lagrangian 
approach. Some of the symmetries present in the theory 
(\ref{SUSYML})
at the classical level are anomalous. It was suggested long ago
 that simple Lagrangians for some effective fields
can summarize all information on the anomalous Ward identities.

Thus, in pure (non-supersymmetric) Yang-Mills theory the trace of 
the energy-momentum 
tensor $\theta_{\mu\nu}$ has anomaly. Correspondingly, one can 
write a Lagrangian for the dilaton field (interpolating the operator of 
the trace of the energy-momentum tensor) which codes all $n$-point 
functions
implied by this anomaly \cite{SM}. 
In supersymmetric gluodynamics the anomalous operators
are $\theta_\mu^\mu$, $\gamma^\mu S_\mu$ and $\partial_\mu
J^\mu$ where $S_\mu$ is the supercurrent and $J^\mu$
is the gluino current. They form a supermultiplet.
The Lagrangian realizing the anomalous Ward
identities can be  naturally constructed \cite{VY} in terms of 
the chiral superfield 
\beq
S = \frac{3}{32\pi^2}W^2 \equiv \frac{3}{32\pi^2}\,\mbox{Tr}\,W^2
\eeq
where
$$
W_\alpha (x_L,\theta ) \equiv\frac{1}{8}{\bar D}^2
\left({\rm e}^{-V} D_\alpha {\rm e}^{V}\right)\, ,
$$
and the color trace above is in the fundamental representation.
 The lowest
component of the superfield $-W^2$ is $\lambda\lambda$
while the $F$ component is nothing but the original
SUSY Yang-Mills Lagrangian, $G^2 +iG\tilde G +i
\lambda^\dagger_{\dot\alpha}D^{\dot\alpha\beta}
\lambda_\beta$. The construction was carried out \footnote{The 
vacuum angle $\vartheta$ in these works was set equal to zero, and 
the $\vartheta$ dependence was not discussed.}
 in Ref. 
\cite{VY} (see also \cite{TVY}); the corresponding    Lagrangian is
\beq
{\cal L} =\left. \left( \bar S S \right)^{1/3}\right|_D + \left( \left.
\frac{1}{3}S\ln ({S^N}/{\sigma^N})\right|_F + \, \mbox{h.c.}\right) 
\, ,
\label{VYL}
\eeq
where 
$\sigma$ is a numerical parameter,
$$
\sigma = {\rm e}\Lambda^3 {\rm e}^{i\vartheta /N}\, ,
$$
$\Lambda$ is the  scale parameter, a positive number of dimension 
of mass which we will set equal to unity in the following.
Finally,  $\vartheta$ is the vacuum angle. Other
 numerical constants irrelevant for our purposes are 
set equal to unity.  

The derivation of Eq. (\ref{VYL}) is pretty straightforward.
The kinetic term is obviously invariant under
the scale transformations and the $R$ rotations,
\beq
S \ra S e^{2i\beta}\, , \,\,\, \theta \ra \theta e^{i\beta}\, .
\eeq
The potential term is not invariant, however. For instance, under
the $R$ rotations,
\beq
\delta {\cal L} \propto \beta \left( \int d^2\theta S - \int d^2\bar 
\theta 
\bar S\right)\, ,
\eeq
which is exactly the anomalous Ward identity for the chiral 
rotations. All other anomalies are then automatically reproduced 
because of the supersymmetry of the VY 
Lagrangian.

After an appropriate rescaling, making the kinetic term
canonical, one gets 
\beq
{\cal L} =\left. \left( \bar \Phi \Phi \right)\right|_D +
\left( \left.
\Phi^3 \ln {\Phi ^N}\right|_F + \, 
\mbox{h.c.}\right) 
,\,\,\,
S = \Phi^3 \, .
\label{VYLR}
\eeq

Up to a redefenition of the superfield this seems to be the only 
Lagrangian which faithfully represents the anomalous Ward 
identities.

This topic was in a dormant state for over a decade. The interest to
this approach was revived recently in connection with the so called
``integrating in" procedure in supersymmetric gauge theories with 
matter (see e.g. Ref. \cite{IS}).

A remark is in order here to explain in what sense 
Eq. (\ref{VYL}) is an effective Lagrangian. 
Clearly it is not
a genuine low-energy effective Lagrangian in the Wilsonian sense.
There are no Goldstone bosons corresponding to the anomalous 
symmetries, and apart from $W^2$, other fields in the theory 
may interpolate
particles with masses of the same order of magnitude as the ones 
retained 
in Eq. (\ref{VYL}). This Lagrangian, therefore, does not arise 
after integrating out of the heavy modes. Rather it is
an effective Lagrangian in the sense that it is  a generating functional 
for
vertex functions of the field components of $W^2$. Of course,
Eq. (\ref{VYL}) should then be understood only as the first two
terms in the derivative expansion. The higher derivative corrections 
to this
expression should be generically large, and therefore Eq. (\ref{VYL})
can not be expected to give a reasonable approximation to
the (on shell) particle interaction vertices.
The effective potential part of this Lagrangian (namely the 
Lagrangian 
evaluated on constant field configurations), however,
should be exact since it is 
determined unambiguously by all  anomalous Ward identities of 
the theory (it includes all relevant $n$-point functions evaluated at 
zero momenta).
Therefore, the Lagrangian is suitable for
 examining the vacuum states of the theory. 

In fact, the last remark requires some qualification.
Although the Lagrangian (\ref{VYL}) has some appealing features,
even a brief examination shows that it can not be complete.
First, the scalar 
potential
following from Eq. (\ref{VYL}) is not a single-valued function of the 
field. If we start, say, at $S= \Lambda^3 $ and travel continuously in 
the 
complex $S$ plane, the value of the scalar potential at $S= {\rm 
e}^{2\pi i}\Lambda^3$
will be different from that at  $S= \Lambda^3$. This is of course 
unacceptable \footnote{Below the lowest component of $S$ is 
denoted 
by $\phi$. Sometimes, when no confusion can arise,  we will still use 
the letter $S$ for the lowest component of superfield.}.
Second, the discrete $Z_{2N}$ symmetry inherent to the original 
theory 
(\ref{SUSYML}) is not reflected in (\ref{VYL}). 
These unsatisfactory features 
were pointed out, e.g. in Ref. \cite{Vene}. 

Our task here is to provide a natural modification to 
this
Lagrangian, which will cure these two problems, while 
preserving the 
correct transformation properties under the anomalous symmetries.

The key observation  is as follows. Since $S$ is supposed to be 
equivalent
to $W^2$, it must satisfy a global constraint ensuring that 
the integral $(1/32\pi^2)\int d^4xG\tilde G$ (in the Euclidean space)
can only take  integer values. This constraint can be 
imposed in an explicitly supersymmetric manner at the Lagrangian 
level by
introducing an integer-valued Lagrange multiplier variable $n$
\beq
{\cal L} =\left. \left( \bar S S \right)^{1/3}\right|_D + 
\left( \left.
\frac{1}{3}S\ln ({S^N}/{\sigma^N})\right|_F + \, \mbox{h.c.}\right) 
+\left.\frac{2\pi i n}{3}\left( S-\bar S\right)\right|_F\, .
\label{newL}
\eeq
Note that the variable $n$ takes only integer values and is not a local
field. It does not depend on the space-time coordinates and, 
therefore,
integration over it 
\footnote{More exactly, one must sum over $n$ in the partition 
function. Similar integer-valued Lagrange multiplier appears in the 
bosonized version of 
the Schwinger model \cite{Smilga}.}
imposes only a global constraint on the 
topological 
charge. It is easy to see that (after the Euclidean rotation) the 
constraint does indeed take the form
$$
\frac{1}{32\pi^2}\int d^4xG\tilde G =Z\, .
$$

Alternatively, one can say that, in calculating
the correlation functions through the  functional integral
with the action (\ref{VYL}), one must sum over all branches of the 
logarithm.
It is perfectly clear that all anomalous Ward identities are kept 
intact. Moreover, this 
 prescription naturally restores the equivalence of all branches 
of the logarithm lost in the original construction \cite{VY}.
The modification suggested is crucial.

The extra term we have
added to the Lagrangian is clearly supersymmetric  and is also 
invariant
under all global symmetries of the original theory. Now  both the 
single-valuedness of the potential and the $Z_N$ 
invariance
are restored. The chiral phase rotation by the angle $2\pi k/N$ with 
integer
$k$ just leads to the shift of $n$ by $k$ units. Since $n$ is summed 
over in the functional integral,
the resulting Lagrangian for $S$ is indeed $Z_N$ 
invariant  \footnote{The 
explicit invariance here 
is $Z_N$ rather than the complete $Z_{2N}$ of the original
SUSY gluodynamics, since we have chosen to write our effective 
Lagrangian for 
the superfield which is invariant under $\lambda\rightarrow 
-\lambda$.}.

Information we are interested in is contained in 
the 
scalar potential that follows from the  effective Lagrangian 
(\ref{newL}), 
since
it is the minima of the scalar potential that determine the
vacua of the theory.
It is instructive to see how the change we propose is reflected in the 
scalar
potential. 
The chiral superfield $S$ for the purpose of calculating the effective 
potential
can be written as
\beq
S=\phi+\theta^2 \frac{1}{\sqrt{2}} (A+{i}B)\, .
\eeq
For the spatially-constant fields the Euclidean action
takes the form
\beq
{\cal A}_E =
\left\{ -\frac{1}{9} (\phi\phi^*)^{-2/3}\frac{1}{2} (A^2-B^2) 
-\frac{N}{3}\sqrt{2} A\ln |\phi | +\frac{iN}{3}\sqrt{2} B\alpha\right\} 
V
\label{euclac}
\eeq
where $\alpha =$ Arg$\phi$ and the quantization condition
enforced by the summation over $n$ in Eq. (\ref{newL}) is
\beq
\frac{\sqrt{2}}{3}BV = b\, , \,\,\,  b = \mbox{integer}\, .
\label{qc}
\eeq
Here $V$ is the full space-time volume and we have set the vacuum
angle $\vartheta = 0$ for the time being. 
If the quantization condition is ignored, elimination of the auxiliary 
fields $A$ and $B$ leads to the 
 original VY  scalar potential 
\beq
U(\phi)=N^2(\phi^*\phi)^{2/3}\ln \phi \ln\phi^{*}
\equiv N^2(\phi^*\phi)^{2/3} \left( \ln^2| \phi |
+\alpha^2\right) 
\, .
\label{oldp}
\eeq
As was mentioned, the result is neither single-valued nor
has it correct periodicity in $\alpha$.
To calculate the corrected effective potential we have to 
take into account the quantization condition (\ref{qc}).
The variable $A$ is unconstrained and can be integrated over in the 
usual way, and
for the variable $B$ integration should be substituted by
summation over integers,
$$
\int dB \ra  \frac{3}{\sqrt{2}}V^{-1} \sum_{b=0,\pm 1,...}\, .
$$
After  the  field $A$ is eliminated, as before, we obtain 
the following 
expression
for the effective potential
\beq
U(\phi)=-V^{-1}\ln \left[\sum_{b=0,\pm 1,  ...}\exp \left\{ 
-VN^2(\phi^*\phi)^{2/3}  \ln^2| \phi |-
\frac{1}{4}(\phi^*\phi)^{-2/3}\frac{b^2}{V} -iN\alpha b 
\right\}\right]
\label{newp}
\eeq

We pause  here to discuss some general features of the effective 
potential (\ref{newp}).
If $b$ could be considered as a continuous variable 
and the summation over $b$
could be
replaced by  integration, we would recover the old potential of 
Eq. (\ref{oldp}).
For small phase angles, $\alpha\ll 1/N$, this is a valid 
approximation, since the coefficients of $b^2$ and $b$ terms are 
small, and the exponent is a 
function
of $b$ which varies very slowly. Therefore, in the vicinity of the real 
axis, the new effective potential is close to 
 the old one. In fact,  if $\alpha = 0$, the corrected potential 
 coincides exactly with that of Veneziano and Yankielowicz. 
It has a minimum at $\phi=1$. However, unlike the 
Veneziano-Yankielowicz potential, $Z_N$ invariance of (\ref{newp})
is explicit: all points $\alpha= 2\pi k/N$ are obviously equivalent, 
and,  in 
particular, there are
 $N$ degenerate minima at
\beq
\phi=e^{i2\pi\frac{k}{N}}\, , \,\,\,  k=0, 1, ... , N-1\, .
\label{minima}
\eeq
This means that away from the line $\alpha  = 0$, the potential 
(\ref{oldp}) gets corrections.  

Expanding the exponent in 
(\ref{newp}) at small $\alpha$ it is possible to conclude
that the small-$\alpha$ expansion of $U(|\phi |, \alpha )$ coincides 
with (\ref{oldp}).
This means that at small $\alpha$
$$
U(|\phi |, \alpha ) \propto \left[ \ln^2 |\phi | +\alpha^2 +{\cal O}(
\exp (-C/\alpha^2))\right]\, .
$$

Consider now the derivative of the potential with respect to 
$\alpha$  at
fixed value $|\phi |$ and $\alpha\neq 0$, 
\beq
\frac{\partial U}{\partial \alpha }=
 iNV^{-1}\frac{\sum_{b=0,\pm 1,  ...}b\, \exp \left\{ 
-VN^2(\phi^*\phi)^{2/3}  \ln^2| \phi |-
\frac{1}{4}(\phi^*\phi)^{-2/3}\frac{b^2}{V} -iN\alpha b 
\right\}}
{\sum_{b=0,\pm 1,  ...}\exp \left\{ 
-VN^2(\phi^*\phi)^{2/3}  \ln^2| \phi |-
\frac{1}{4}(\phi^*\phi)^{-2/3}\frac{b^2}{V} -iN\alpha b 
\right\}}\, .
\label{deriv}
\eeq
This expression has the meaning of the average density of 
instantons minus the average density of anti-instantons in the 
Yang-Mills 
theory on the real axis  but with the shifted  
 value of the vacuum angle $\vartheta= -N\alpha$ (see Eq. 
(\ref{uvt}) below).
Clearly, this expression is finite for any value of $\alpha$. The 
effective 
potential is therefore a continuos function of the field. 
Consider now the rays $\alpha=\frac{\pi}{N}k$ where $k$ is odd. 
These rays are exactly in the middle between the ``valleys"
$\alpha=\frac{2\pi}{N}k$.
For these values of the angle the weight in the sum over $b$ in the 
numerator of Eq. (\ref{deriv}) is 
symmetric
under $b\rightarrow -b$. The average in the numerator of Eq. 
(\ref{deriv}),  therefore, 
vanishes
\footnote{There can be no ``spontaneous breaking'' of the symmetry
$b\rightarrow -b$
 since Eq. (\ref{deriv}) is a simple sum over $b$ rather than a
functional 
integral.}. 
Thus,  along the directions Arg$\phi=\frac{\pi}{N}k$ where $k$ is 
odd,  the 
effective
potential has the topography of a ridge.

We conclude this section by briefly discussing  the 
$\vartheta$ dependence. From Eq. (\ref{VYL}) it is clear
that $\vartheta$ enters in the scalar potential only through the 
combination
\beq
U_\vartheta (|\phi |, \alpha ) = U(|\phi |, \alpha 
-\frac{\vartheta}{N})\, . 
\label{uvt}
\eeq
When $\vartheta$ continuously varies from 0 to $2\pi$, the 
``mountain ridge" picture rotates by $2\pi/N$: the first valley 
becomes the second, and so on, cyclically.  Such a picture was 
predicted from a general consideration \cite{SHVA}.

\section{The Vacuum State without Gluino Condensate}

The  following feature of the
scalar potential (\ref{newp}) is important for our consideration. In 
addition  to $N$ minima of Eq. (\ref{minima}) it exhibits 
an 
unexpected 
solution at $\phi = 0$. To reveal the extra solution 
it may be convenient to proceed from the superfield 
$S$ to the superfield $\Phi$, whose kinetic term has the canonical 
form. These superfields are related, $S =\Phi^3$; the same relation 
holds for the lowest components, $\phi =\varphi^3$ where $\varphi$ 
is the lowest component of $\Phi$.
The zero $\langle\Phi\rangle$  solution corresponds to the vanishing 
gluino condensate, and its interpretation has been never  discussed 
previously. 
This zero energy state at $\langle\Phi\rangle = 0$
reflects
a phase of the supersymmetric gluodynamics
with no breaking of the  $Z_{2N}$ symmetry and vanishing gluino 
condensate. 

The occurrence of the condensate-free phase may sound suspicious, 
since 
superficially 
this statement contradicts the Witten-index 
argument \cite{WI}. Indeed, the Witten index for the $SU(N)$ group 
is $N$,
which is exactly equal to the number of the 
$\langle\lambda\lambda\rangle \neq 0$ states of vanishing energy
one obtains from Eq. (\ref{minima}).  
So, the only way to reconcile the existence of 
the extra state at $\phi = 0$ with this result
is that it should not contribute to the Witten 
index. 

Surprisingly, it is very difficult to rule out this possibility,
and  this may indeed be the case.
For this to happen there must exist  an equal number of $F$ = even
and $F$ = odd states at $\Phi =0$. The Lagrangian (\ref{newL}) does 
imply the existence of the massless fermion mode in the 
condensate-free regime. 
Usually, in the 
Wess-Zumino type models one can always introduce the mass term 
to the chiral superfield considered, all massless modes are 
eliminated, and the zero-energy states concentrated near the zeros of 
the superpotential are all of the bosonic type. In which case,  they 
certainly 
contribute to the Witten index. That is not true for 
the effective Lagrangian (\ref{newL}). Here its form is rigid,
since it merely reflects the anomalous Ward identities as well as the 
discrete
non-anomalous symmetries of the
underlying theory. The mass term is forbidden --
it would explicitly violate the Ward identities, and
the vacuum structure obtained in this way will have nothing to do 
with that of the
underlying SUSY gluodynamics. For instance, it would break 
explicitly the 
$Z_{2N}$ symmetry and eliminate all  $Z_{2N}$ breaking vacua  in 
Eq. (\ref{minima}).

If the excitation modes  are strictly massless, in general
it is very difficult to decide which state is $F$-even and which is
$F$-odd in the case of the unbroken supersymmetry, when the 
supercharge acts trivially on the vacuum. Therefore on the basis 
of the effective potential alone we are unable to determine what is
the contribution of the $\phi=0$ states to the Witten index.
We will argue later that, in fact,  the $(-1)^F$ counting of the 
zero-energy states may depend 
strongly
on the nature of higher derivative terms in the effective action, 
which
we have neglected so far and which are not determined by
the anomalous
Ward identities.

Having argued that the  problem with the Witten index need 
not be an obstruction, let 
 us present now a positive, although  subtle,  argument in favor of 
the
existence of the additional  $\langle\lambda\lambda\rangle = 0$ 
vacuum state in SUSY gluodynamics. To this end we need to make a 
digression and recall   a puzzle
with the dynamical calculation of the gluino condensate.

Calculation of the gluino condensate \cite{NSVZ1}  was the first 
application of instantons 
in  supersymmetric  gluodynamics in the strong coupling regime. 
Consider for simplicity the
$SU(2)$ gluodynamics. In this case there are four gluino zero modes 
in the
instanton field and, hence, there is no direct instanton contribution to 
the 
gluino condensate $\langle \lambda\lambda \rangle$. At the same 
time the instanton does contribute to the 
correlation 
function
\begin{equation}
\langle \lambda^a_\alpha (x)\lambda^{a\alpha}(x)
,\lambda^b_\beta (0)\lambda^{b\beta}(0) \rangle\, ,
\label{I1}
\end{equation} 
Here $a,b =1,2,3$ are 
the color
indices and $\alpha ,\beta = 1,2$ are the spinor ones. An explicit  
instanton calculation \cite{NSVZ1} shows that the correlation 
function (\ref{I1}) is equal to a non-vanishing constant.

At first sight this result might seem like supersymmetry-breaking 
since
the instanton does not generate any boson analog of Eq. (\ref{I1}).
Supersymmetry does not forbid, however, a non-vanishing result for 
Eq. 
(\ref{I1}) provided that this two-point function is actually an
$x$ independent constant. 

Three elements are crucial for the proof of the above 
assertion: (i) the supercharge 
${\bar Q}^{\dot\beta}$ acting on the vacuum state annihilates it; 
(ii) ${\bar Q}^{\dot\beta}$ commutes with $\lambda\lambda$; 
(iii) the derivative $\partial_{\alpha\dot\beta}(\lambda\lambda)$ is 
representable as the anticommutator of ${\bar Q}^{\dot\beta}$ and 
$\lambda^\beta G_{\beta\alpha}$. The second and the third point 
follow from the fact that 
$\lambda\lambda$ is the lowest component of the chiral superfield 
$W^2$, while $\lambda^\beta G_{\beta\alpha}$ is its middle 
component.

Now, we differentiate Eq. (\ref{I1}),  substitute 
$\partial_{\alpha\dot\beta}(\lambda\lambda)$ by $\{
{\bar Q}^{\dot\beta}, \lambda^\beta G_{\beta\alpha}\}$ and obtain 
zero.  Thus,  supersymmetry requires the $x$ derivative of (\ref{I1}) 
to 
vanish \cite{NSVZ1}. 
This is exactly what happens if the correlator (\ref{I1}) is  a constant.
		
If so, one can compute the result at short
distances where it is presumably saturated by small-size instantons, 
and, 
then, the very same constant is predicted at large distances, 
$x\rightarrow
\infty$. On the other hand, due to the cluster decomposition 
property,
which must be valid in any reasonable theory, the correlation 
function (\ref{I1})
at $x\rightarrow
\infty$
reduces to $\langle\lambda\lambda \rangle^2$. Extracting the 
square root we 
arrive at a
(double-valued) prediction for the gluino condensate. 

Instead of exploiting strong coupling
calculations (which is always suspect), one can  use a fully 
controllable approach in the weak 
coupling regime \cite{SHVA}. Namely, one  extends the theory by 
adding one flavor (two chiral matter superfields, doublets with 
respect to the gauge $SU(2)$), carries out  the calculation in the 
weakly coupled Higgs 
phase (i.e. assuming a large value of the Higgs field), and then 
returns back to SUSY gluodynamics by exploiting 
the holomorphy of the condensate in the mass parameter 
\cite{SHVA}.

The result for $\langle\lambda\lambda\rangle$
obtained in the strong coupling regime 
(i.e. by following the program outlined after Eq. (\ref{I1})) does {\em 
not} match
$\langle\lambda\lambda\rangle$ calculated in this  indirect but  
fully controllable way. As a matter of fact, as was shown in Ref. 
\cite{NSVZ},
\beq
\langle\lambda\lambda\rangle_{\rm scr}^2 =
\frac{4}{5} \langle\lambda\lambda\rangle_{\rm wcr}^2\, ,
\label{DISCR}
\eeq
where the subscripts scr and wcr mark the strong and weak coupling 
regime calculations. Since the weak coupling regime calculation
seems to be flawless, 
suspicion naturally falls on the strong coupling analysis.
But where concretely is the loophole? 

A tentative  answer might be found in the hypothesis put forward by 
Amati {\em et al}. 
\cite{Amati}. It was assumed that, instead of providing us with the 
expectation value of 
$\lambda\lambda $ in the given vacuum, instantons in the strong 
coupling regime  yield
an average value of $\langle\lambda\lambda \rangle$ in all possible 
vacuum 
states. In the weak coupling regime, we have a marker: a large
classical VEV of the Higgs field tells us in what particular  vacuum 
we 
do our instanton calculation. In the strong coupling regime, such a
marker is absent. 

This hypothesis by itself, however, does not explain the discrepancy 
 (\ref{DISCR}), if there are only two vacua characterized by 
$\langle\lambda\lambda \rangle =\pm \Lambda^3$.
The gluino condensate is not affected by 
the averaging over these two vacuum states,
since the contributions of these two vacua to Eq. (\ref{I1}) are equal. 
If, however, there 
exist 
extra zero-energy states with $\langle\lambda\lambda \rangle = 0$ 
which are involved in the averaging, the final result in the strong 
coupling regime
is naturally different from that obtained in the weak coupling regime
in the {\em given } vacuum. Moreover, the value of the condensate 
calculated in the strong coupling approach should be smaller, 
consistently
with Eq. (\ref{DISCR}).

Note that the approach of Ref. \cite{SHVA} has nothing to say about 
possible states at the origin. If there is a minimum at the vanishing  
expectation value of the Higgs field (and vanishing gluino 
condensate) the theory is always in the strong coupling regime near 
this minimum. Introduction of the matter fields does not help to
push the theory in the weak coupling regime. 

Now it is in order to return to the Witten-index argument,
in a bid to use it in a positive aspect. 
The existence of extra vacua can 
potentially
resolve another long-standing paradox in SUSY gauge theories.
It has been known for a long time that in $SO(M)$ SUSY gauge 
theories the
Witten index does not coincide with
the number of different  $\langle\lambda\lambda\rangle \neq 0$ 
states of vanishing energy
one obtains from the instanton calculations
in the weak coupling regime \cite{SHVA}. The former is rank +1
while the latter is $M-2$ for the orthogonal groups. 
The anomalous and 
non-anomalous
symmetry structure of these models is similar to that of $SU(N)$ 
gauge
theories. The effective potential would, therefore, have the same
form as Eq. (\ref{newp}) with the  parameter $N$ substituted by $M-
2$. It would, 
therefore, still exhibit a minimum at $\phi=0$. It is possible that for 
orthogonal groups the number of fermionic states at this point is 
larger
than the number of bosonic ones, and the difference precisely
makes up for the difference between the Witten index (rank +1) and 
the 
number of $Z_{M-2}$-breaking bosonic minima with the 
non-vanishing gluino condensate.

\section{Dynamical Consequences}

In this section we present some speculations as to the 
nature of
the $Z_{2N}$-symmetric vacuum. First, the energy of this state 
vanishes
and therefore supersymmetry  is unbroken. Second, the effective 
Lagrangian 
indicates
the existence of massless fermions. This, as noted before, is a 
necessary
condition to avoid the contradiction with the Witten-index 
calculation.
The scalar field $\varphi$ also is massless near this point, and 
obviously 
is
the superpartner of the massless fermion. 
The existence of strictly massless particles in this phase of 
supersymmetric gluodynamics is remarkable since they are not 
Goldstone modes associated with the spontaneous breaking of some 
symmetry. As a matter of fact, no symmetry is spontaneously 
broken, and still the massless modes are present.
The situation is somewhat reminiscent of conformally invariant 
phases
of some other SUSY gauge theories (with matter) which figure so 
prominently
in recent work on electric-magnetic duality \cite{ISR}. In fact the 
similarity 
may
well be even closer.

Taken at its face value, the
effective potential (\ref{newp}) would lead one to conclude that 
the
scalar quartic self-coupling near $\varphi=0$ diverges 
logarithmically
at zero momentum.
However, as pointed out earlier,  one can not use this effective 
potential
to reliably establish interactions between the particles. This is 
especially
true near the point $\varphi=0$. The reason is obvious.  For 
non-constant
fields one has to take into account higher derivative terms in the 
effective
action. The anomalous
Ward identities, while unable to determine these
higher derivative terms completely, do impose some restrictions on 
their form.
In particular, to preserve the Ward identity following from the 
dilatational
symmetry, every extra derivative should be accompanied roughly by 
the factor
$(\bar S S)^{-1/6}$.
Example of this type of terms is
\beq
\frac{(\partial_\mu \bar S^{1/3}\partial_\mu S^{1/3})}{(\bar 
SS)^{1/3}}\, .
\eeq
Near the point $\phi=0$, starting already at this low order in
derivatives, the corrections explode. The derivative expansion is not 
valid
even for very low momenta.
This suggests that the Green's functions of the fields $S$ and $\bar S$
will have non-analytic behavior in  momentum. With the 
knowledge of the existence of massless excitations, we are lead to 
conjecture that at $\varphi=0$ at low momenta, the theory is 
conformal. 

The precise nature of this conformal theory (anomalous dimensions, 
etc.)
can not be determined on the basis of the effective potential alone.
It is natural to expect that  counting of the number of fermionic 
and
bosonic vacua at zero would  depend heavily on the properties of the
conformal theory.
For example, for $SO(N)$ and $SU(N-2)$ gauge theories, the effective 
potentials would be the same, but the conformal theories at zero 
could
be completely different. That could explain the difference in the 
number 
of the fermionic and bosonic vacua in the two cases needed to 
comply 
with the
Witten-index calculation. 

If the chirally invariant phase of pure 
gluodynamics
does indeed exist, it has drastic consequences for 
supersymmetric theories with massless (or light) matter.
Indeed,  at $N_f< N_c - 1$,  the Affleck-Dine-Seiberg (ADS) 
superpotential \cite{ADS} (see also Ref. \cite{TVY}; $M_i^j$ is an 
$N_f\times N_f$ moduli 
matrix)
\beq
{\cal W} \propto \left(\frac{\Lambda^{3N_c-N_f}}{\mbox{det} 
M}\right)^{\frac{1}{N_c-N_f}}
\label{ADSp}
\eeq 
is generated through the gluino condensation in the unbroken
strongly coupled $SU(N_c-N_f)$ gluodynamics. 
If the latter has the condensate-free
phase, the original SUSY QCD will have a 
branch with no superpotential generated. Previously, a similar 
phenomenon (an additional branch with no superpotential)
was observed \cite{INSE} in the $SO(N)$ theories with $N_f= N-4$.
For $ M_i^j\neq 0$ the gauge symmetry is broken down to
$SU(2)\times SU(2)$. It was noted \cite{INSE} that the gaugino
condensates corresponding to different $SU(2)$'s can have
opposite signs, so that the sum vanishes, implying vanishing 
superpotential. 

The existence of two inequivalent branches can be argued
from a different perspective.  Assume we introduce the mass term
\beq
\Delta{\cal L}_{\rm tree} = m \mbox{Tr}\, M|_F
\label{mmt}
\eeq
to  the superpotential of the theory.  For simplicity,  the mass 
parameters for all flavors are  equal. Then the matter mass 
term 
breaks
$U(1)_R$ explicitly, but the discrete $Z_{2N}$ remains unbroken
at the Lagrangian level. When $m$ is very large, we return back to 
supersymmetric gluodynamics, with the solutions
\footnote{Note that at large $m$ the combination 
$m^{N_f/N_c}\Lambda^{(3N_c-N_f)/N_f}$ is a proper scale parameter 
of supersymmetric gluodynamics.}
\beq
\langle\lambda\lambda\rangle \sim m^{N_f/N_c}\Lambda^{(3N_c-
N_f)/N_f}\, , \,\,\,  \langle M_i^j \rangle \sim \delta_i^j m^{(N_f-
N_c)/N_c}\Lambda^{(3N_c-N_f)/N_f}\, ,
\label{c1}
\eeq
for the first branch and 
\beq
\langle\lambda\lambda\rangle  = 0\, , \,\,\, \langle M_i^j \rangle = 
0\, ,
\label{c2}
\eeq
for the second. 
The functional $m$ dependence  is actually known exactly,
for all $m$
\cite{SHVA} (see also \cite{Sei}), due to its holomorphic nature.
This allows one to analytically continue the results to small $m$.
Two solutions indicated in Eqs. (\ref{c1}) and (\ref{c2})
correspond to two different branches of SUSY QCD. 

In the massless limit, i.e. $m\ra 0$, at the origin 
($\langle\lambda\lambda\rangle = M_i^j =0$), the theory possesses 
an 
unbroken
chiral symmetry, the $R$ symmetry. Therefore, the anomalous AVV 
triangles must be matched -- they must be the same  at
the fundamental and composite levels \cite{GHooft}.
The matching implies the existence of a rich spectrum of massless 
baryons, which goes far beyond one massless fermion residing
in the superfield $W^2$ in the condensate-free phase of 
supersymmetric gluodynamics. Another possibility is that the 
$R$-symmetry is spontaneously broken, implying the existence of 
the massless ``$\eta '$" and its fermion superpartner. The problem 
with the latter scenario is 
that no appropriate  order parameter for the $R$ symmetry breaking 
can be immediately found, since the obvious candidates, 
$\langle\lambda\lambda\rangle$ and $M_i^j$ , vanish.  It is possible, 
that a non-chiral field condenses, for instance the lowest component 
of ${\overline{W}}^2 M_i^j$. This expectation value would break both, 
the 
$R$ symmetry and the axial $SU(N_f)$. 

If $N_f= N_c - 1$, the ADS superpotential is generated by instantons. 
If det$M\neq 0$ one can carry out calculation of the superpotential 
in the 
weak coupling regime, where the result is unambiguous (and 
coincides, of course, with that of Affleck {\em et al.}).
In this way one arrives at the standard picture of a 
run-away vacuum, with the zero energy state at det$M =\infty$.
The superpotential, however, is not defined at det$M= 0$. The 
condensate-free phase of supersymmetric gluodynamics 
implies
a supersymmetric vacuum solution of SUSY QCD at
det$M= 0$.

Finally, at $N_f= N_c $ the superpotential is not generated.
A continuous manifold of degenerate vacua exists \cite{Nati},
\beq
\mbox{det} M - B\tilde B = \Lambda^{2N}\, .
\label{vacman}
\eeq
At large det$M$ the gauge symmetry is completely broken, the 
theory is in the weak coupling regime, the instanton calculation
is unambiguous, and one can explicitly check, by doing 
the instanton calculation, that the quantum moduli space is indeed 
described by Eq. (\ref{vacman}).  We suggest that there is an extra 
point in the vacuum manifold, characterized by
\beq
\mbox{det} M = 0, \,\,\,  B=\tilde B = 0\, ,
\label{extrapoint}
\eeq
that cannot be continuously reached from the weak coupling regime.

To illustrate this, we again switch on the matter mass term
(\ref{mmt}), with the intention of continuing analytically  from the 
limit of the large mass (pure gluodynamics) to the limit of the zero 
mass (massless SUSY QCD). The holomorphy in mass implies in this 
theory that the gluino condensate is proportional to $m$ while
$M_i^j$ and $B,\tilde B$ are proportional to $m^0$. 
As a result,  the conventional Seiberg solution
corresponds to the point $B=\tilde B = 0$ and  det$M=\Lambda^{2N}$ 
on the manifold (\ref{vacman}). 
If a branch of gluodynamics with the vanishing gluino condensate
exists, then 
$B=\tilde B = 0$ and  det$M =0$ on this branch. Continuing 
analytically 
in $m$ to the massless limit we observe the extra point 
(\ref{extrapoint})
on the vacuum manifold, disconnected from Seiberg's solution. 
 
\section{Conclusions}

An elegant picture of supersymmetric gauge dynamics
which was gradually evolving in the eighties and took a much more 
complete form after a recent breakthrough \cite{Nati},
is still not free from question marks. We suggest
an unorthodox solution eliminating all of them, completely.
The key element of our consideration is the existence
of the condensate-free phase of supersymmetric gluodynamics, with 
the
unbroken $Z_{2T(G)}$ symmetry. This additional vacuum is 
supersymmetric, and can be compatible with other known aspects of 
SUSY gauge dynamics only provided that  massless excitation modes, 
bosonic and fermionic, are present  in this vacuum. Corresponding 
modifications must also take place in the theories with the massless 
matter (SUSY QCD). 

Although our conjecture seems compelling, keeping in mind the
non-trivial  nature of our consideration, which cannot be tested in 
the weak coupling regime, it is natural to be cautious. Independent 
confirmation and a more thorough understanding of  the dynamics of  
unconventional
massless bound states characteristic of the condensate-free phase 
is highly desirable. In particular, it is desirable to learn how to 
calculate the number of the $F$-even and $F$-odd zero-energy states
in the extra minimum. A simpler task would be to detect a mismatch 
between the strong-coupling and weak-coupling calculations for
gauge groups other than $SU(2)$. Studying this mismatch as a 
function of the group constants may reveal a pattern of ``leakage"
in the $Z_{2T(G)}$ unbroken phase. 

\vspace{0.3cm}

{\bf Acknowledgments}: \hspace{0.2cm} 

One of the authors (M.S.) is grateful to G. Veneziano  for  stimulating 
discussions of the VY effective Lagrangian. We would like to thank
I. Kogan, S. Rudaz, A. Smilga, M. Voloshin  and S. Yankielowicz for 
useful comments.

This work was supported in part by DOE under the grant number
DE-FG02-94ER40823.

\end{document}